\documentclass[journal]{IEEEtran}

\usepackage{authblk}
\usepackage{cite}
\usepackage{amsmath,amssymb,amsfonts}
\usepackage{algorithmic}
\usepackage{graphicx}
\usepackage{textcomp}
\usepackage[table,xcdraw]{xcolor}
\usepackage[breaklinks=true,colorlinks,citecolor=blue,linkcolor=blue,urlcolor=blue]{hyperref}

\newcommand*{\hermconj}{^{\mathsf{H}}}

\title{On the Use of Singular Value Decomposition as a Clutter Filter for Ultrasound Flow Imaging}
\title{On the Use of Singular Value Decomposition as a Clutter Filter for Ultrasound Flow Imaging}
\author[1]{Kai Riemer}
\author[1]{Marcelo Lerendegui}
\author[1]{Matthieu Toulemonde}
\author[2]{Jiaqi Zhu}
\author[3]{Christopher Dunsby}
\author[1]{Peter~D.~Weinberg}
\author[1]{Meng-Xing Tang}
\affil[1]{Department of Bioengineering, Imperial College London, London, United Kingdom}
\affil[2]{Department of Mechanical Engineering, University College London, London, United Kingdom}
\affil[3]{Department of Physics, Imperial College London, London, United Kingdom}
\date{April 2023}

\begin{document}

\maketitle

\begin{abstract}
Filtering based on Singular Value Decomposition (SVD) provides substantial separation of clutter, flow and noise in high frame rate ultrasound flow imaging. The use of SVD as a clutter filter has greatly improved techniques such as vector flow imaging, functional ultrasound and super-resolution ultrasound localization microscopy. The removal of clutter and noise relies on the assumption that tissue, flow and noise are each represented by different subsets of singular values, so that their signals are uncorrelated and lay on orthogonal sub-spaces. This assumption fails in the presence of tissue motion, for near-wall or microvascular flow, and can be influenced by an incorrect choice of singular value thresholds. Consequently, separation of flow, clutter and noise is imperfect, which can lead to image artefacts not present in the original data. Temporal and spatial fluctuation in intensity are the commonest artefacts, which vary in appearance and strengths. Ghosting and splitting artefacts are observed in the microvasculature where the flow signal is sparsely distributed. Singular value threshold selection, tissue motion, frame rate, flow signal amplitude and acquisition length affect the prevalence of these artefacts. Understanding what causes artefacts due to SVD clutter and noise removal is necessary for their interpretation.

\end{abstract}


\section{Introduction}
\IEEEPARstart {D}{oppler} ultrasound has been widely used in clinical practice for the visualization and quantification of blood flow in arteries and veins \cite{Hoskins1999}. The technique assumes that blood and tissue signals are separable in the Doppler spectrum, which implies that blood must flow faster than any tissue movement. This assumption holds true in large arteries, away from their walls, but starts to fail for near-wall or microvascular flow. Microbubble contrast agents enable specific imaging of blood flow even with very slow or no flow, through nonlinear pulsing sequences such as pulse inversion or amplitude modulation. However, as the ultrasound frequency increases beyond the resonant frequency of the population of microbubbles, nonlinear imaging becomes less effective \cite{Tang2006, Yildiz2015, Zhu2019}. To increase the flow signal amplitude and to suppress tissue and noise signals, digital clutter filter techniques are used. These techniques are often applicable both in conventional and contrast enhanced imaging.   

Singular value decomposition (SVD) is a powerful matrix factorization method that has been widely applied in ultrasound research \cite{Fort1995, Bjaerum2002, Kruse2002, Mamou2009, Demene2015, Arnal2017}. SVD has been used to suppress noise and tissue clutter particularly for flow imaging at high ultrasound frequencies and imaging frame rates \cite{Geunyong2014, Errico2015, Errico2016, Lin2017a}. However, just as classical filtering can introduce ringing artefacts, SVD clutter filtering can introduce changes in image intensity over space and time \cite{Alberti2016, Song2017, Baranger2018, Riemer2021}. Consequently, misinterpretation of the image e.g., in quantification of perfusion, vessel-wall segmentation or localization may occur. A blinking of the microbubble signal after SVD based clutter filtering but not before \cite{Brown2019} and changes in intensity of the flow signal based on the blood flow velocity have been reported \cite{Riemer2021}. Such errors may be clinically important: for example, intensity changes observed in perfusion protocols are used to distinguish benign and malignant masses \cite{Kong2014}. And the estimation of carotid flow velocity and wall shear stress may be affected, which are important in the investigation of cardiovascular disease \cite{Likittanasombut2006, Riemer2020}.

The reasons why SVD as a clutter filter generates artefacts include 1) rank deficiency and its limitations as a linear operation that may not effectively remove non-linear clutter, e.g. nonlinear propagation, 2) its inability to handle changes in the clutter distribution over time, e.g. motion, 3) its assumptions about the data, e.g. slow flow, and 4) its dependence on the proper selection of parameters, such as the number of singular values to retain \cite{Wahyulaksana2022, Ozgun2021}. 

The aim of this study is to demonstrate and define the artefacts that appear due to SVD as a clutter filter in two-dimensional ultrasound imaging of flow, and to provide a guidance on what to consider when using SVD. A combination of conventional and contrast-enhanced images are used to showcase artefacts. These examples include how the threshold, flow velocity or signal amplitude not only influence the signal to noise ratio (SNR) in filtered images but also the prevalence of artefacts. For illustrative purposes, examples are selected that maximize artefacts, and improper use of SVD is deliberate to show artefacts occurring due to both approximation errors and incorrect thresholds. 

First, the basic concept of SVD clutter filtering is described. Next, examples of the artefacts in simulations and in experiments are demonstrated and artefacts are categorized. It is shown that the user's choice of a threshold can vary based on the selection method and that tissue motion can cause a persistent artefact. Last, methods for mitigating artefacts such as motion correction (MoCo) and a sliding window SVD filter are examined, the results are discussed and the principle for the cause of the artefacts is explained. 

\section{Methods}

\subsection{Singular Value Decomposition}
Ultrasound flow imaging data $s(x,z,t) \in \mathbb{C}$ can be described by three signal components: high amplitude tissue and clutter $c(x,z,t)$, fast moving blood or microbubbles $b(x,z,t)$ and Gaussian random noise $n(x,z,t)$, where $t$ is the time of acquisition and $x$ and $z$ are the lateral and axial coordinates. The most common method to perform SVD on two-dimensional ultrasound images is to reorganize the signal $s(x,z,t)$ into a two dimensional spatiotemporal Casorati matrix $S \in \mathbb{C}^{n_x n_z \times n_t}$, where each column describes a single instant in time, e.g. a frame, and each row a single pixel over time \cite{Demene2015}. Here, $(n_x,n_z)$ is the size of each frame and $(n_t)$ the number of frames. The Casorati matrix is factorized by SVD into a weighted, ordered sum of separable matrices.

\begin{equation} 
 S=UDV\hermconj
\end{equation}

The columns of $U$ are the left singular vectors and corresponding columns of $V$ are the associated right singular vectors. Here, the symbol $V\hermconj$ represents the Hermitian transpose of the matrix $V$. $U$ and $V$  are orthonormal and also known as the spatial and temporal singular vectors respectively. D is a diagonal matrix, where diagonal elements are the singular values which are ordered in size and decrease as the column number increases. Mathematically, the decomposition corresponds to finding the eigenvectors and eigenvalues of the covariance matrices $SS\hermconj$ and $S\hermconj S$.   

\begin{equation} 
 SS\hermconj=UDD\hermconj U\hermconj
\end{equation}
\begin{equation} 
 S\hermconj S=VD\hermconj DV\hermconj 
\end{equation}

The eigenvectors form an orthonormal basis for describing the spatial and temporal variance of the data. The eigenvectors can be thought of as a least squares linear fit of the direction of maximal residual variance and the eigenvalues give a weighting to how large a contribution each signal component makes to the whole data set. In other words, SVD identifies signal components in each individual pixel with similar characteristics over space and time. These signal components are broken down into corresponding spatial and temporal eigenvectors and eigenvalues. Each spatial vector is a static image that is modulated along time by the complex signal of its corresponding temporal eigenvector and weighted by its singular value. Each combination of corresponding singular vectors and values creates a sequence of a static image with changing intensity. The image dimensions $n_z \cdot n_x$ and the number of frames $n_t$ determine the row-rank of the singular value matrix. The row-rank determines the number of singular values and vectors and, consequently, how sensitive of a threshold selection can be made. 

\subsection{Singular Value Filtering}
SVD filtering relies on the assumption that clutter, blood and noise are each represented by different subsets of singular values; they must be uncorrelated and lay in orthogonal subspaces. Eigenvectors corresponding to low singular values are typically associated with low velocities and a low spread of Doppler frequencies. This is because tissue consistently moves together over time and thus exhibits high spatiotemporal coherence. Only a low number of eigenvalues are required to represent tissue and clutter, despite the energy, which is often orders of magnitude stronger than that of the flow. Blood flow has less spatiotemporal coherence and spreads over a large number of singular vectors. The subspace defined by the eigenvectors corresponding to the lowest singular values is associated with randomly distributed noise. Note, however, that noise is spread everywhere in the spectrum. By combining singular values a filtered dataset is reconstructed

\begin{equation} 
 S_{filtered}=\sum_{l=i}^{u} U_{l}D_{l,l}V_{l}\hermconj
\end{equation}

Normally, filtering is performed by defining two cut off points in the spectrum: a high magnitude, low index $i$ is selected to exclude tissue signals whose singular value magnitude is greater than $i$, and a lower magnitude, higher index $u$ is selected to remove noise, by excluding all singular values below the magnitude of $u$. The reconstructed data uses all singular values whose magnitude is between $i$ and $u$.

\subsection{SVD cut-off point selection} \label{methods_cut_off}
The selection of $i$ and $u$ is subject to the user’s choice. In practise the described subspaces do not always separate neatly. It is important to systematically determine the cut-off points to minimize misassignment. A fixed threshold based on a priori knowledge might not be optimal. Three methods of threshold selection have proven to result in a good suppression of tissue signal. The first (U-method) is based on spatial similarity of spatial singular vectors \cite{Baranger2018}. This method assumes separability in space by correlating the spatial vectors and defining a subdomain in the shape of a square in the correlation matrix. The second method (S-method) uses the turning point and point of linear decrease of the singular value magnitude curve as the tissue and noise threshold, respectively. In the third method (V-method), the first threshold is marked by the first inflexion point and the second threshold by the plateau of the singular vector Doppler curve \cite{Song2016, Waraich2019}. This method assumes separability based on temporal characteristics with tissue motion and noise having set frequencies. For continuity in subsequent figures the thresholds were selected based on the U-method.  

\begin{center}
    \begin{table*}[h]
    \centering
    \caption{Overview of Simulation and Imaging Parameter of Datasets}
    \begin{tabular}{c|c|c|c|c|c|c|c|c}
    \textbf{Figure} & \textbf{Name}  & \textbf{TX Freq. [MHz]} & \textbf{Frame rate [Hz]} 
    & \textbf{Num. angles} & \textbf{Num. frames} & \textbf{Contrast} & \textbf{Motion} & \textbf{Type} \\ \hline
    \rowcolor[HTML]{EFEFEF}
    1 & Three dots        & n.a.  & n.a.      & n.a.   & 100      & n & n & Animation \\
    2 & Microvasculature    & 7.24  & 100       & 3      & 100      & y & y & Simulation \\
    \rowcolor[HTML]{EFEFEF} 
    3 & Tube                & 5     & 100       & 10     & 500      & y & y & In vitro \\
    3 & Kidney 1            & 4.5    & 1,000    & 5      & 1,500    & y & n & Rabbit \\
    \rowcolor[HTML]{EFEFEF} 
    4 & Capillary           & 4     & 1,000     & 5      & 1,000    & y & n & In vitro \\
    5 & Aorta 1             & 9     & 1,500     & 5      & 1,500    & y & y & Rabbit \\
    \rowcolor[HTML]{EFEFEF} 
    5 & Tumour              & 18    & 600       & 15     & 900      & y & y & Mouse \\
    6 & Kidney 2            & 4     & 500       & 5      & 1,530    & y & n & Rabbit \\
    \rowcolor[HTML]{EFEFEF} 
    7 & Aorta 2             & 9     & 1,500     & 5      & 1,500    & n & y & Rabbit \\
    7 & Aorta 3             & 9     & 1,500     & 5      & 1,500    & y & y & Rabbit \\
    \end{tabular}
    \label{tab:ov1}
    \end{table*}
\end{center}

\subsection{Example datasets and imaging platform}
Two simulations, two \textit{in vitro} acquisitions and six \textit{in vivo} datasets obtained by high frame rate, plane wave imaging with and without contrast agent are presented (Table \ref{tab:ov1}). The first simulation data was created without ultrasound physics. The second was simulated with BUbble Flow Field (BUFF, \cite{Lerendegui2022}). The \textit{in vitro} and \textit{in vivo} imaging was performed with a Verasonics Vantage 256/128/64 LE research ultrasound system with L11-4v, L22-18 and GEL3-12D probes. The transmit frequency was between 4-18 MHz and the Mechanical Index $<$ 0.15. Images were acquired continuously or intermittent. For the contrast-enhanced imaging Sonovue (Bracco, Milan, Italiy) or in-house produced microbubbles were injected by continuous infusion or as a bolus via the appropriate paths. The in-house manufactured microbubbles were lipid-shell filled with decafluorobutane \cite{Lin2017b}. The SVD clutter filter was applied post-acquisition to the beamformed data after Hilbert transformation. All post-processing and image visualization steps were performed in Matlab (MathWorks, USA) including MoCo using \textit{imregdemons}. For SVD the \textit{'econ'} option was selected. For this analysis the contrast to tissue ratio (CTR) is defined as $20 \cdot log_{10}(signal/reference)$. 

\subsection{\textit{In silico} examples}

\subsubsection{Moving dots}
Three dots with constant intensity move from left to right. There were two static tissue regions, one above and one below the region where the dots move each covering 20\% of the image. The length of the sequence was 100 frames. The image dimensions are 512 by 512 pixel. The dots are 40 pixel in diameter and move 4, 2, 1 pixel per frame. Their starting pixel positions are 181, 256, 331 in the y-direction and 56 in the x-direction.   

\subsubsection{\textit{In silico} microbubble flow}
A random microvascular flow network was generated using BUFF \cite{Lerendegui2022} with microbubbles between 0-10 $\mu m$ radius (mode of 2.5 $\mu m$), flowing at a velocity between 0 and 60 mm/s. On the side was a mirrored C-shaped region of tissue. To mimic motion the region of interest (ROI) was sheared after image formation as described by Equation \ref{eq:trans} using the Matlab function \textit{imwarp}, where $n$ is the number of the current frame

\begin{equation} 
A(n) = 
\begin{pmatrix}
1 & 0 & 0 \\
n\cdot e^{-4} & 1 & 0 \\
0 & 0 & 1\\
\end{pmatrix}
\label{eq:trans}
\end{equation}

Plane wave imaging, at a pulse repetition frequency of 300 Hz with three angles (frame rate 100 Hz), was simulated to mimic the bandwidth and geometry of a linear array L11-4v probe at a pulse transmit frequency of 7.24 MHz. White Gaussian beamformed noise with a SNR of 20 dB was introduced for a total number of 100 frames. 

\subsection{\textit{In vitro} examples}

\subsubsection{\textit{In vitro} small vessel imaging}
Two vessel-mimicking phantoms were made, one with a sparse and one with a dense scatter distribution. The first consisted of a 0.4 mm (internal) and 0.8 mm (external) tube (PY2 72-0191 PolyE Polyethylene Tubing, Biochrom, UK) inside a tissue-mimicking phantom of 10\% gelatin (G2500, Sigma Aldrich), 0.3\%-potassium sorbate (85520, Sigma Aldrich), and 89.7\%-water (percentage of total weight). The second phantom consisted of a 300 $\mu m$ capillary (Platinum Cured Silicone Tube, Havard Apparatus, UK) inside agar containing glass beads (45-90 $\mu m$ diameter, Potters Ballotini Ltd, UK). Flowing Sonovue microbubbles with a concentration of 7.5 $\cdot$ 10\textsuperscript{4} mb/mL or 2.5 $\cdot$ 10\textsuperscript{7} mb/mL were administered using a programmable syringe pump (PHD 2000, Harvard Apparatus, UK). The pump was connected to one end of the tube and produced mean flow velocities of 5 mm/s and 4, 10 and 80 mm/s. In the first setup transmitted pulses were centered at 5 MHz and the frame rate was 100 Hz for a contrast imaging sequence with two apertures and 10 angles acquired with the multiplexed GEL3-12D probe. The radio frequency data were beamformed to yield 500 frames of compounded images. To have natural motion the probe was handheld while imaging. In the second setup a linear probe L12-3v and 5-angle compounding plane wave imaging was used. Transmitted pulses were centered at 4 MHz and the frame rate was 1000 Hz. The radio frequency data were beamformed to yield 1000 frames of compounded images. The change in CTR of different ensemble lengths was studied. A ROI was selected resembling the shape of the tube in the vessel phantom. The ROI for background tissue was set to be 3 mm above the ROI for the tube. 

\subsection{\textit{In vivo} examples}

All animal experiments were authorised by the Animal Welfare and Ethical Review Body of Imperial College London and followed the Animals (Scientific Procedures) Act 1986. 

\subsubsection{Mouse tumor} 
A mouse tumor model (female, 7 weeks, Charles River UK Ltd, UK) was imaged at a frame rate of 600 Hz, with 15 angles yielding a pulse repetition frequency of 9 kHz. The transmit frequency was 18 MHz with a single half cycle transmission. The total number of frames was 900. The transducer was fixed by a probe holder. 

\subsubsection{Rabbit aorta}
High frame rate plane wave imaging of the abdominal aorta of a New Zealand White rabbit (male, 11 weeks, HSDIF strain, Specific Pathogen Free, Envigo, UK) was performed. The imaging frame rate was 1500 Hz with a total of 5 angles. The transmission frequency was 9 MHz with two half cycles for a total duration of 1 s corresponding to 1500 frames. The probe was fixed during the acquisition. 

\subsubsection{Rabbit kidney}
The kidney of NZW rabbits (male, 10-12 weeks, HSDIF strain, Specific Pathogen Free, Envigo, UK) was investigated with plane wave images acquired at a pulse repetition frequency between 2.5 - 5 kHz adding up to a total imaging time between 1.5 - 3 s. Two different imaging schemes were tested. A continuous plane wave contrast enhanced imaging sequence with five angles spanning 12$^\circ$ transmitted at a frame rate of 1 kHz and an intermittent acquisition acquired at 0.5 kHz. The centre frequency was 4.5 or 4 MHz with two half cycles. The total number of frames was 1500 or 1530. The probe was handheld during the first acquisition and fixed with a probe holder for the second acquisition.  

\section{Results}

\subsection{Types of artefacts generated by SVD clutter filtering}

\begin{figure}[h]
\centering
\includegraphics[width=.49\textwidth]{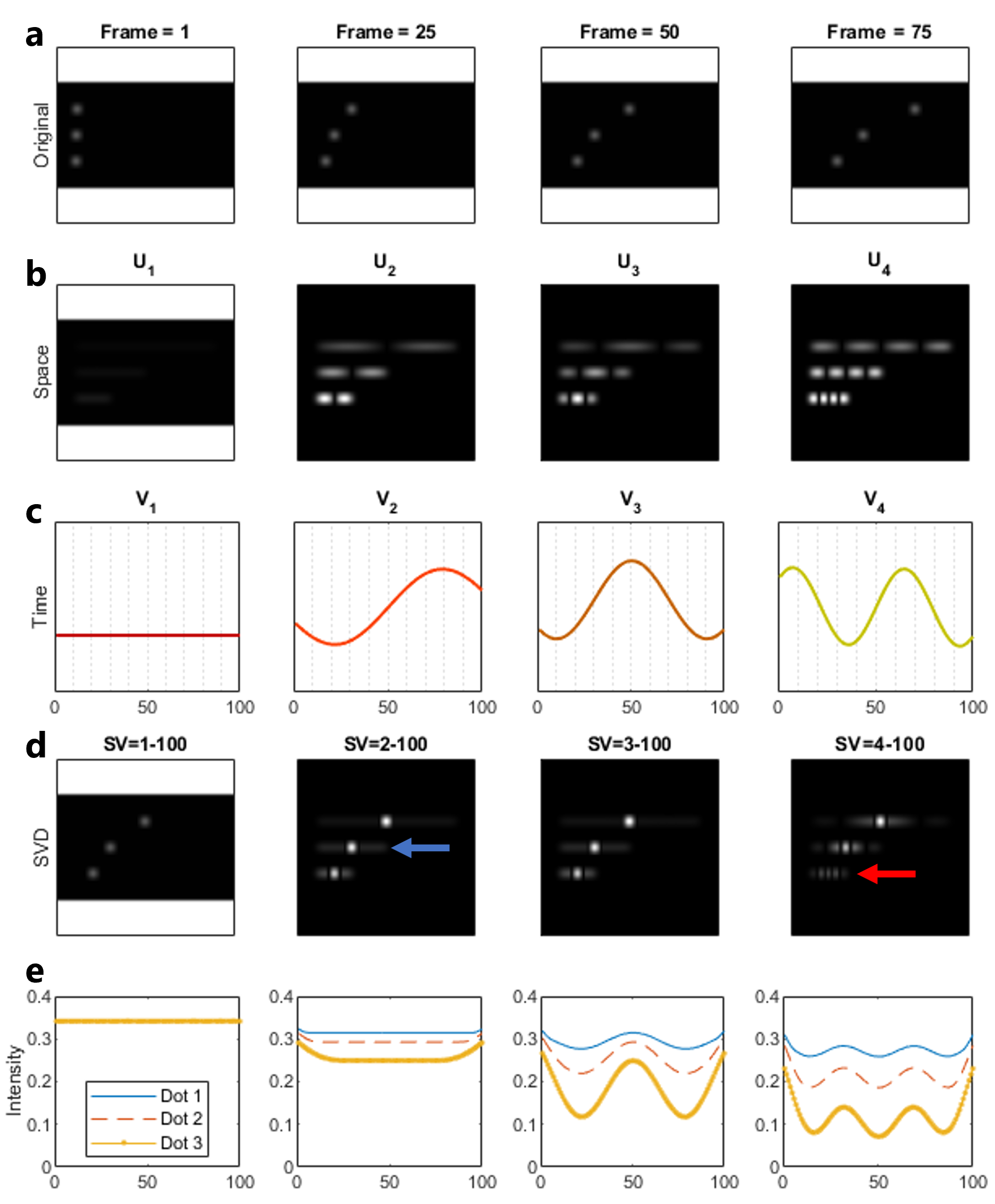}
\caption{Three types of artefact generated by SVD clutter filtering. (a) Three dots move from left to right. A single frame from a 100 frame sequence of images is shown for each. The white blocks on the top and the bottom mimic tissue. (b) and (c) show the first four spatial and temporal representations of the acquisition. The 50th frame of the reconstructed sequence when removing none, the first, up to the second or up to the third singular value is presented (d). The intensity of each dot over all frames is shown in (e). Flashing (marked in plots of e), ghosting (marked by blue arrow) and signal splitting (marked by red arrow) are observed. Images are normalized and their magnitude is displayed with a dynamic range from 0 to 1.}
\label{fig01}
\end{figure}

Fig. \ref{fig01} demonstrates the three types of artefact - namely flashing, ghosting and splitting. Flashing was defined as temporal intermittent intensity variation, ghosting was defined as an image duplication or smearing and splitting was defined as a partitioning or widening of the signal. These artefacts are not present in the original data (Fig. \ref{fig01} a) but they exist in the clutter filtered data (Fig. \ref{fig01} d). The left (spatial) and right (temporal) singular vectors for each image are shown in Fig. \ref{fig01} b and c, respectively. Fig. \ref{fig01} d represents the final reconstructed image at frame 50 and  Fig. \ref{fig01} e shows the intensity of each dot over all frames. With increasing singular value the spatial and temporal frequencies increase (Fig. \ref{fig01} b and c). As a result, local and global intensity changes occur, which can be perceived as flashing (Fig. \ref{fig01} e), ghosting (marked by blue arrow) or splitting (marked by red arrow). The artefacts differ for the three dots travelling at different velocities.  

\subsection{SVD artefacts in ultrasound images}
\begin{figure}[h]
\centering
\includegraphics[width=.48\textwidth]{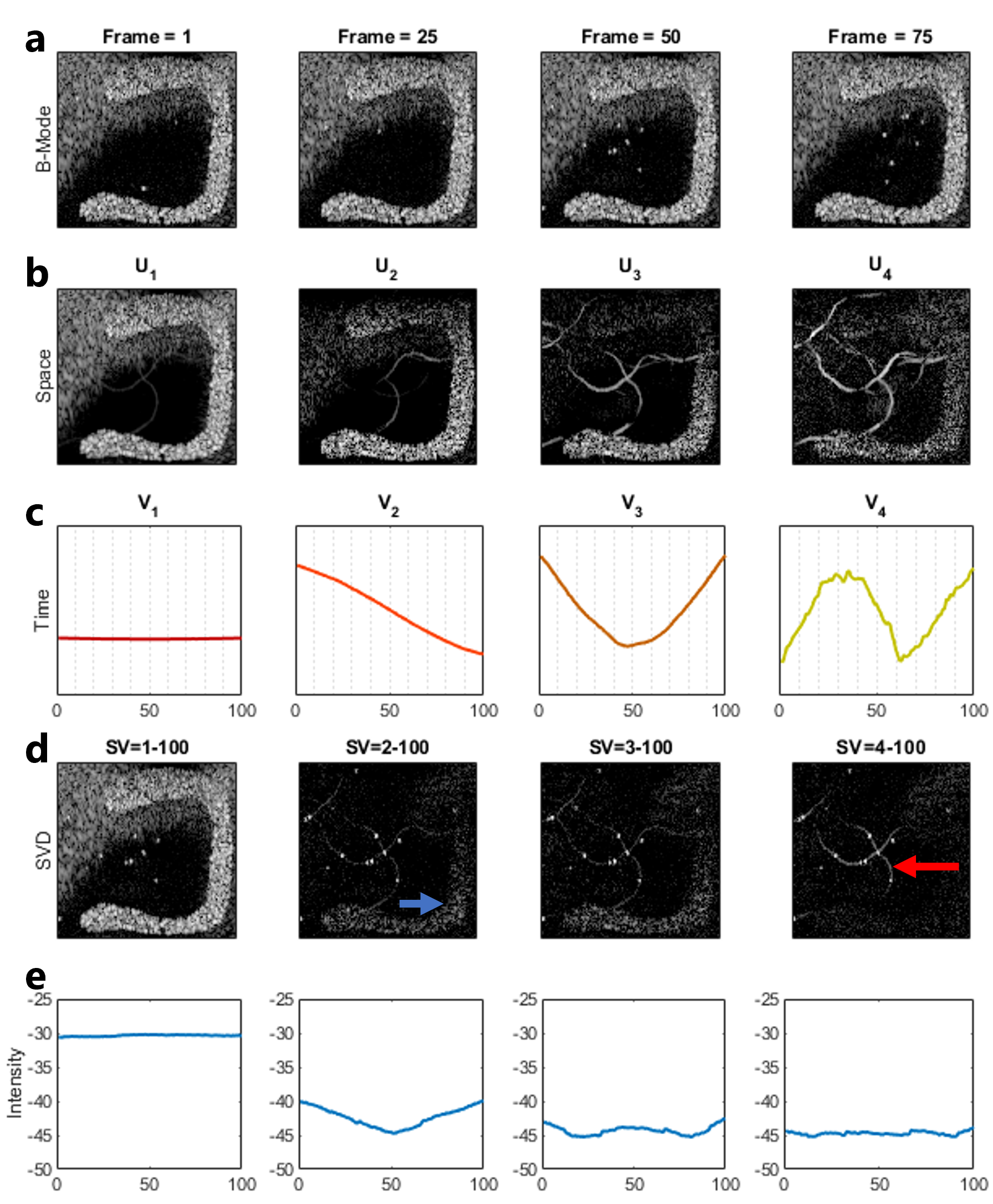}
\caption{Local and global changes of signal intensity in an \textit{in silico} model of the vasculature with motion. (a) Illustrates the log compressed B-Mode data at different time points. The C-shaped region of the right mimics tissue. (b) and (c) show the spatial and temporal vectors of the decomposition and (d) shows the reconstructed partial projection after SVD clutter filtering at frame 50 and (e) shows mean image intensity over time. Total length of acquisition is 100 frames. The shearing motion leads to imperfect separation of the flow. Visible image artefacts appear in (d) marked by blue and red arrow for different lower thresholds e.g. 2, 3 or 4.}
\label{fig02}
\end{figure}

Fig. \ref{fig02} a, b and c show the B-Mode sequence, the corresponding left (spatial) and right (temporal) singular vectors of a simulated flow structure. Fig. \ref{fig02} d represents the reconstructed images at frame 50 for four different cut-off points of the 100 frame long sequence and Fig. \ref{fig02} e shows the mean image intensity over all frames. The example resembles that of Fig. \ref{fig01} but is from simulated radio frequency data with global motion. Underestimation of the clutter subspace (i$<$4) causes the tissue region marked by blue arrow to flash (Fig. \ref{fig02} e). Over the length of the sequence the intensity of clutter changes. Overestimation (i$=$4) elicits strong ghosting (marked by red arrow). The tissue suppression is motion dependent. The displacement from the shearing transformation is higher with depth and the suppression of clutter is inversely related to the amount of tissue motion. The spatial vectors (Fig. \ref{fig02} b) illustrate a significant overlap between tissue and microbubble; the third and fourth spatial singular vectors contain information from both the clutter and flow domain at a similar intensity. 

\subsection{Singular value threshold selection}
\begin{figure}[h]
\centering
\includegraphics[width=.48\textwidth]{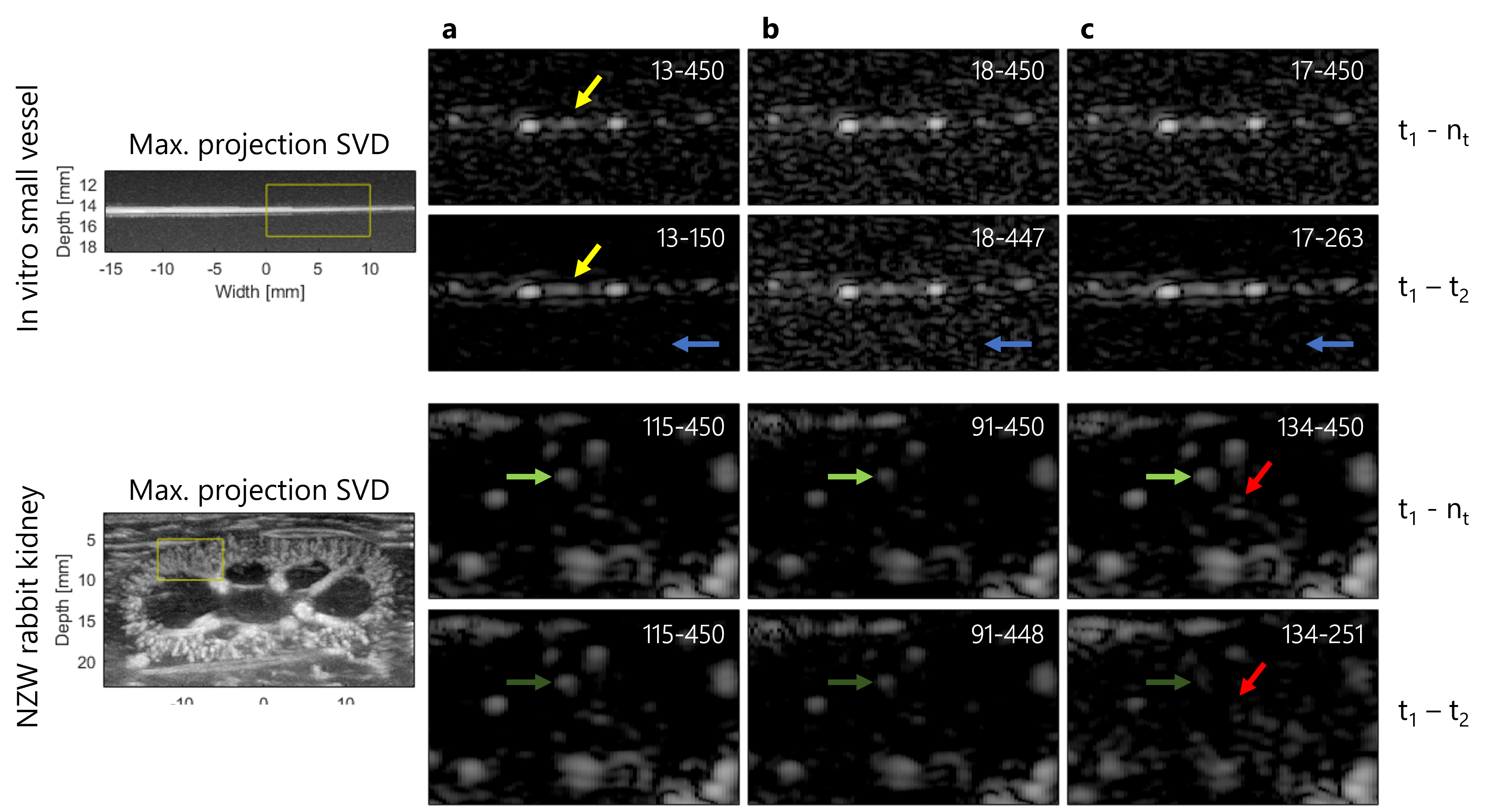}
\caption{Effect and artefacts at different singular value thresholds. Top: Small tube filled with microbubbles and manual acquisition. Bottom: contrast enhanced acquistion of a rabbit kidney with fixed probe position. Singular value cut-off methods: (a) spatial similarity, (b) singular value magnitude and (c) frequency threshold selection method. A small change in threshold selection can have large impact on the appearance of the filtered image. Yellow, green and red arrows indicate regions where local intensity is strongly dependent on threshold. All images are displayed with 40 dB dynamic range.}
\label{fig04}
\end{figure}

The selection of thresholds in SVD clutter filtering is essential but not trivial. Fig. \ref{fig04} qualitatively demonstrates the effect of the threshold when using the most common selection methods (U-, S- and V-method) for singular value threshold selection. Fig. \ref{fig04} top displays a small vessel \textit{in vitro} and Fig. \ref{fig04} bottom shows a contrast enhanced plane wave imaging sequence of a rabbit kidney. Column a, b and c correspond to the three different methods (U-, S-, V-method). Fig. \ref{fig04} $t_1$ - $n_t$ shows results obtained when using only a lower threshold (low matrix index) and $t_1$ - $t_2$ is for a lower plus a higher cut off (high and low matrix index). The thresholds are shown in each image. A small change in cut-off can change the relative intensity distribution. The change is most marked for the regions denoted by the yellow and red arrows. The red arrow shows a microbubble when including all but the first 134 singular values but no bubble is seen when excluding the singular values 252-450.  

\subsection{Stack size, flow speed and flashing}
\begin{figure}[h]
\centering
\includegraphics[width=.48\textwidth]{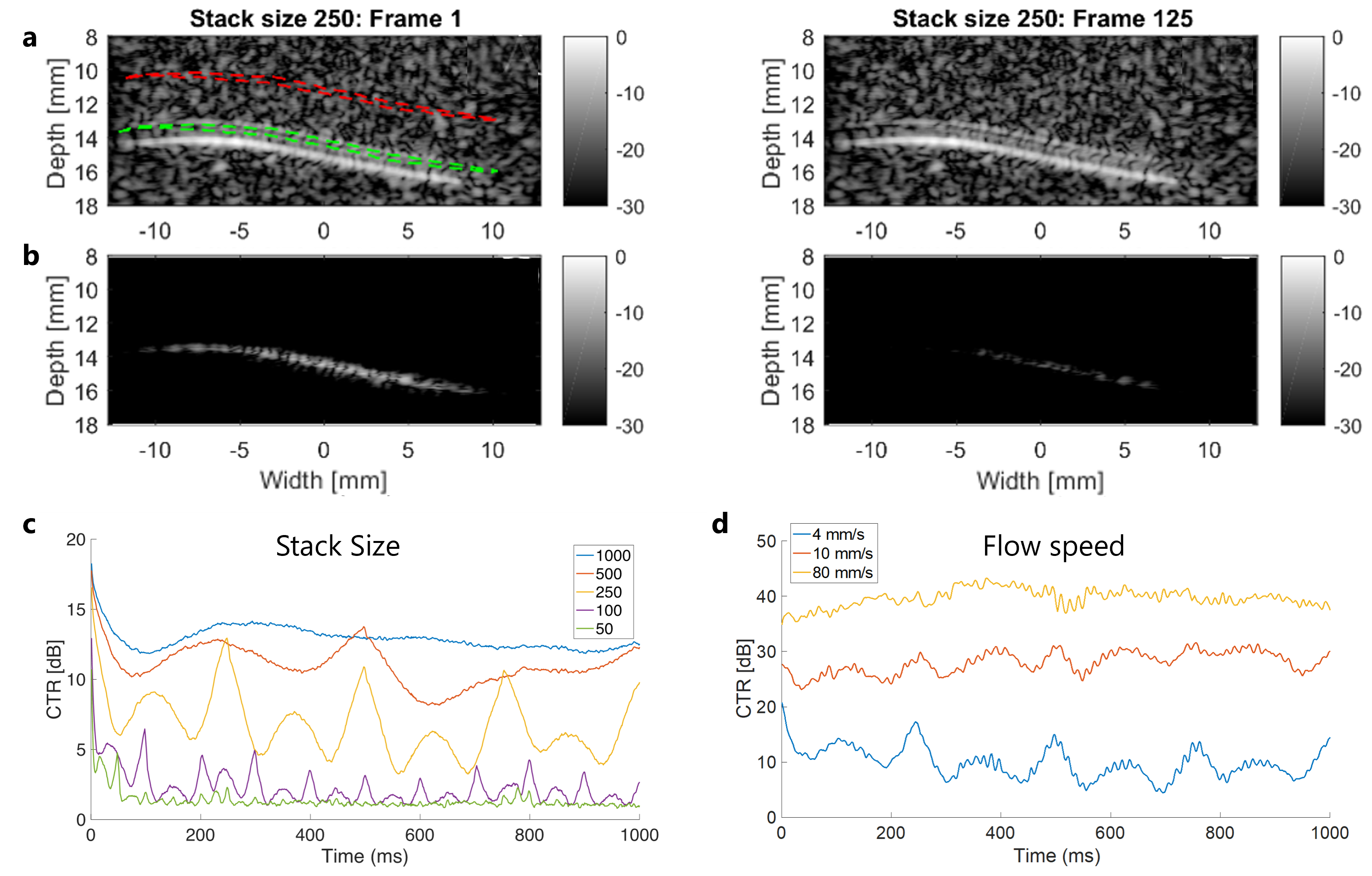}
\caption{Flashing artefact introduced by SVD in an \textit{in vitro} capillary flow phantom. (a) Unfiltered and (b) SVD filtered images of the first and 125th frame of a 250 image stack, respectively. (c) Variation in CTR as a function of stack size with 4 mm/s flow. (d) Variation in CTR as a function of flow speed. The red and green polygon mark the ROIs for the CTR measurement.}
\label{fig05}
\end{figure}

Flow speed and the number of frames influence the variation in intensity over time. Fig. \ref{fig05} shows the flashing artefact in an acquisition of a capillary phantom. Fig. \ref{fig05} a shows the original B-Mode data at frame 1 and 125 of a 250 image stack. Fig. \ref{fig05} b shows the filtered data. Unlike in the previous example microbubbles are more densely distributed. Instead of local or individual variation, the intensity changes globally. The vessel darkens and brightens over the whole width of the image and over time. Intensity changes are dependent both on the number of frames used for SVD (Fig. \ref{fig05} c) and the flow speed (Fig. \ref{fig05} d). A shorter stack size causes higher variation in CTR, and an increase in flow velocity reduces the variation of the CTR. 

\begin{figure*}[h]
\centering
\includegraphics[width=.98\textwidth]
{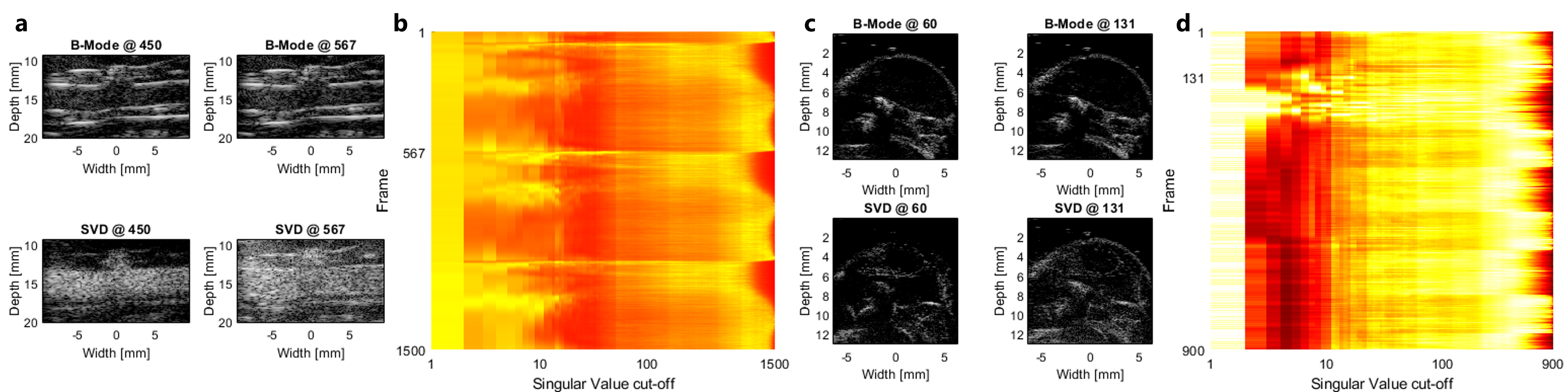}
\caption{Demonstration of the flashing artefact independent of cut-off threshold. (a, c) B-Mode and SVD filtered images of rabbit abdominal aorta and mouse tumor with 40 dB dynamic range at time points Frame = 450 and Frame = 567 during the acquisition (frame rate = 1500). (b, d) Normalized tissue intensity over time and as a function of cut-off point; each column is normalized by itself. In (b) the tissue intensity changed at a rate of 3 Hz regardless of cut-off point. This is not true of the intensity of the unfiltered B-Mode data which stayed constant. The flashing manifested as a cyclic illumination of the whole image to the point where it became hard to distinguish between tissue and flow.}
\label{fig07}
\end{figure*}

\subsection{Tissue motion, threshold and flashing}
Fig. \ref{fig07} demonstrates the flashing artefact independent of cut-off threshold due to motion. In the case where the measurement is defined by the morphological state e.g. wall shear stress estimation the correction of motion may be undesirable. Fig. \ref{fig07} a shows the rabbit aorta near the origin of the renal artery and Fig. \ref{fig07} c shows the tumour in a mouse model. The motion of the aorta is caused by the cardiac rhythm which changes the diameter of the vessel, and by breathing, whereas motion in the mouse appears in the form of irregular twitching and non-rigid deformation. Figures \ref{fig07} b and d presents the normalized tissue intensity over time as a function of cut-off point and time; each column is normalized by itself. Fig. \ref{fig07} a and the first column of Fig. \ref{fig07} b show that in the original B-Mode image the contrast enhanced blood and tissue intensity remains constant throughout the 1 s of acquisition, which is equivalent to 3 cardiac cycles. Conversely, in the SVD filtered dataset the mean intensity of blood and tissue fluctuate. Large  changes of tissue intensity can be observed at frame 567 (0.37 s) which coincide with the expansion of the vessel during systole. The clutter signal is harder to distinguish from the flow signal compared to at frame 450. Similarly, the signal of the skin over the tumour can be suppressed at still-stand (Fig. \ref{fig07} c, frame 60) but during the rapid twitching the skin and clutter is brighter than the microvasculature. Remarkably, Fig. \ref{fig07} b and d illustrate that the flashing is independent of the cut-off threshold. Here, cut-off 1 includes all singular values, cut-off 2 includes all singular values but the first, cut-off 3 all but the first two and so on. In Fig. \ref{fig07} b underestimation of the clutter subspace (i$<$12) causes irregular flashing and both correct estimation and overestimation (i$=>$12) leads to flashing at a rate of 3 Hz. Selecting a different cut-off point, to the right at the cost of SNR and to the left at the cost of reduced clutter removal, does not change the occurrence of the flashing artefact. The same behaviour can be observed in the mouse tumour model Fig. \ref{fig07} d, albeit to a lesser extend. 

\subsection{Motion correction and signal suppression}
\begin{figure*}[h]
\centering
\includegraphics[width=.98\textwidth]{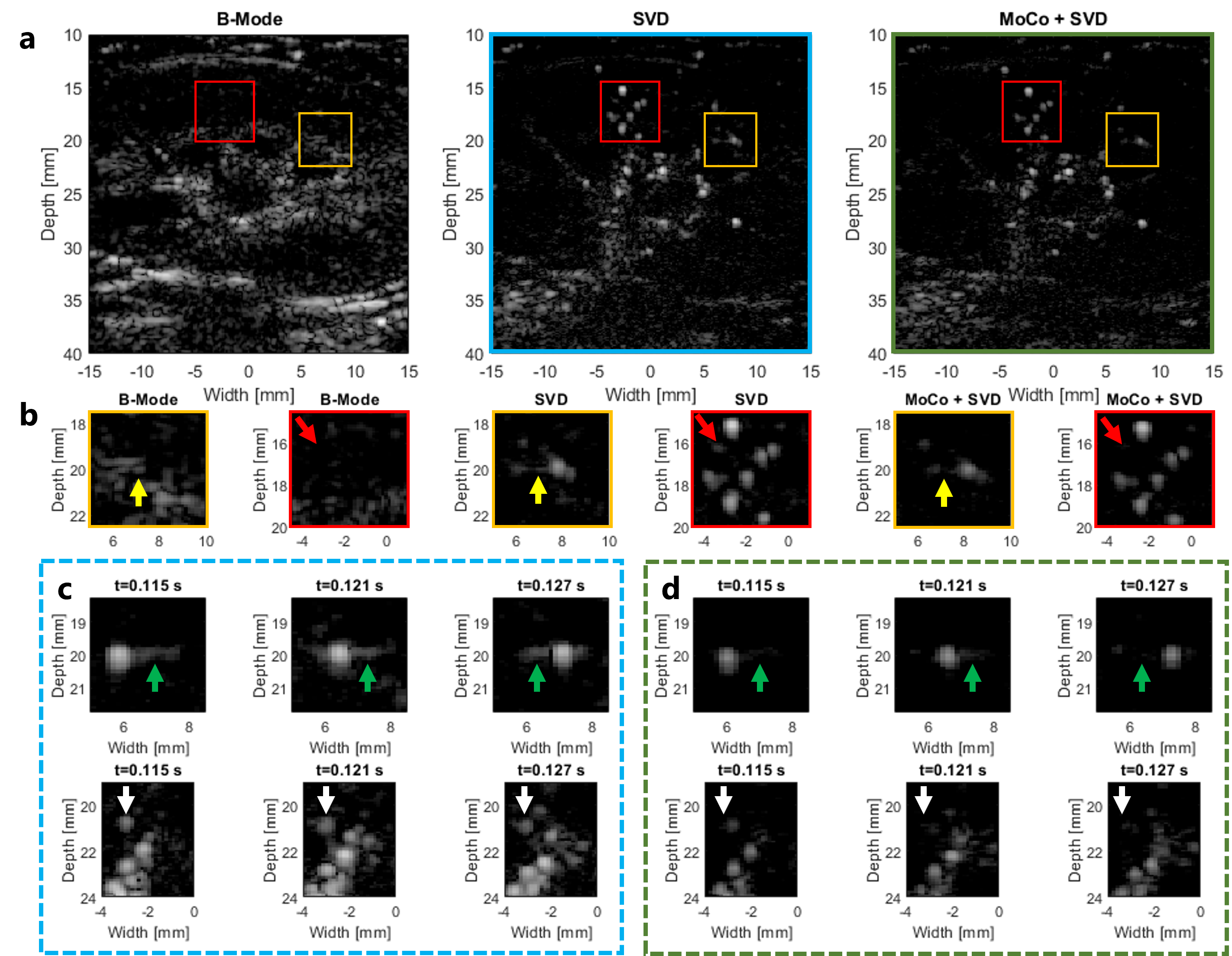}
\caption{Comparison of SVD clutter filtered images without and with prior MoCo. (a) Shows the original B-Mode data, the SVD filtered data and the MoCo+SVD filtered data at a single point in time. (b) Two region of interest which show a ghosting artefact without MoCo (yellow arrow) and suppression of static microbubbles with MoCo (red arrow). (c) Temporal evolution of the ghosting artefact (green arrow) and the suppression of static microbubbles (white arrow). All images are displayed at 40 dB dynamic range.}
\label{fig08}
\end{figure*}

If compensation for tissue motion is feasible its use before SVD clutter filtering might be advantageous. Fig. \ref{fig08} shows a contrast enhanced acquisition of the rabbit kidney with and without motion compensation prior to SVD clutter filtering. The images are displayed at 40 dB dynamic range. Fig. \ref{fig08} b illustrates two zoomed in regions of interest and Fig. \ref{fig08} c and d show the temporal evolution of SVD with and without MoCo, respectively. In the yellow square of Fig. \ref{fig08} b ghosting can be observed without MoCo (yellow arrow), while the red square shows that with MoCo some static microbubbles are suppressed (red arrow). The shape of the signal differs between the two methods. In the top row of Fig. \ref{fig08} c a strong ghosting effect marked by the green arrow can be seen. It is not apparent in the top row of Fig. \ref{fig08} d (even when the dynamic range is extended to 80 dB). The bottom row of Fig. \ref{fig08} c and d again illustrate that MoCo applied before SVD filtering leads to the suppression of the signal of static microbubbles as illustrated by the white arrow. With MoCo fewer microbubbles can be observed and their relative amplitude is changed. 

\subsection{Moving window SVD and lowering CTR}
\begin{figure*}[h]
\centering
\includegraphics[width=.98\textwidth]{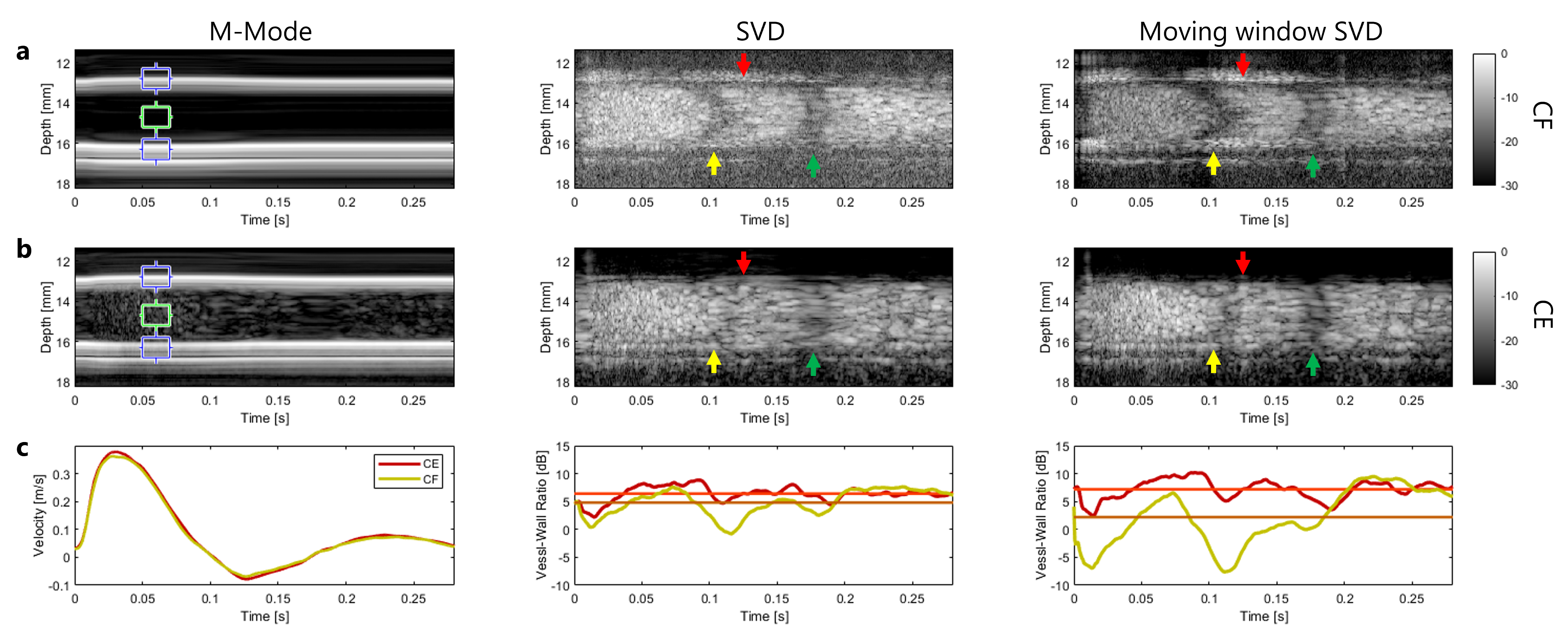}
\caption{Mitigating flashing artefact through sliding window SVD in the rabbit aorta. (a) Conventional contrast agent free acquisition. (b) Contrast enhanced acquisition. (c) Velocity waveform and vessel-wall ratio. Red arrow indicates intensity variations of the wall. Yellow and green arrow show how flow velocity changed the ability of SVD to suppress clutter. The flow signal tapers in region of backward and net zero flow.}
\label{fig09}
\end{figure*}

Tissue motion can also be countered by using fast frame rates and short acquisition times. By using a sliding window SVD filter artefacts due to motion can be mitigated in certain situations. Fig. \ref{fig09} shows the flashing artefact in the rabbit aorta for a conventional (Fig. \ref{fig09} a, CF) and a contrast enhanced (Fig. \ref{fig09} b, CE) acquisition. The first column illustrates the M-Mode from the central region of the image, the second column the M-Mode after SVD filtering and the third column the M-Mode after a moving window SVD filter. The x-axis represents time. Fig. \ref{fig09} c shows the velocity waveform and the vessel-wall ratio in dB. The blue and green ROIs mark the position where the vessel-wall ratio was measured. Applying the SVD clutter filter to the entire length of the contrast enhanced acquisition led to flashing. Dividing the sequence into shorter image stacks using a moving window SVD reduced the flashing of the wall (red arrow) and raised the mean vessel-wall ratio by 0.7 dB (Fig. \ref{fig09} c). However, the lower signal intensity of the blood speckle in the non-contrast acquisition did not benefit from the same sliding window SVD (Fig. \ref{fig09} a). In fact, a sliding window led to a total loss of flow signal (yellow and green arrows). More frame and a higher frame rate are needed. In the contrast-enhanced acquisition, the sliding window SVD introduced changes in intensity of the flow signal of up to 10 dB but reduced the changes in intensity of the tissue region and increased the difference in signal strength between the two. As the flow velocity changed, the ability of SVD to suppress clutter changed too, as illustrated by the tapering of the signal over time (green and yellow arrows).

\section{Discussion}
In this work we have demonstrated artefacts of SVD as a clutter filter in high frame rate blood flow imaging. Namely, we categorized three artefacts: flashing, ghosting and splitting. Flashing was defined as temporal intermittent intensity variation, ghosting was defined as an image duplication or smearing, and splitting was defined as a partitioning or widening of the signal. In the examples flashing manifested itself as local and global changes in intensity over time, ghosting could be observed as a bright patch of signal along the trajectory of a microbubble, and splitting caused a bubble to split into multiple. The examples where chosen to maximize the visual impact of the artefact and may be less apparent in some applications. Users of SVD as a clutter filter should be aware of the artefacts and should consider the impact of the artefact's appearance on their data. 

SVD is a linear operation that transforms the data into a lower-dimensional representation by discarding some singular values. The creation of artefacts in the filtered data was caused because of a loss of information. The loss can be either caused due to an overlap between clutter, flow and noise or an over- or underestimation of the subdomains. For example the flashing artefact is caused by removing singular values corresponding to high energy, e.g. the first few singular values. These correspond to removing a low frequency, high energy sinusoidal component of the signal as illustrated in Fig. \ref{fig01}. By transforming the data into a lower-dimensional representation and discarding parts of it the filtered image has spatial and temporal characteristics complementary to the removed components. The ghosting and splitting artefacts seen in Fig. \ref{fig01} and Fig. \ref{fig02} can be explained similarly. A target’s spatial position is a superposition of different singular vectors and each represents a certain energy and frequency range. The ability to extract flow requires a sufficient number of singular values relative to the distribution of velocities and the frame rate. The number of singular values is determined by the minimum of number of frames and spatial coordinates of the image. S being fully ranked presents the ideal case -- maximum number of singular values with minimum data. In Fig. \ref{fig01} SVD cannot accurately represent the different spatial positions of all targets equally, given the defined ensemble length, frame rate and velocities. This causes ghosting when targets are sparsely distributed. The decomposition into partial descriptions of the signal in space and time can also lead to an incomplete spatial representation of target features when too many high singular values are removed. In Fig. \ref{fig01} d the slowest dot appears to have split into three. With the removal of high singular values, ghosting was first observed, followed by splitting. Slow moving bubbles, which were represented by higher singular values, experience this effect first as shown Fig. \ref{fig01} and Fig. \ref{fig08} with the removal of static bubbles. The fast moving dot in Fig. \ref{fig01} is only exhibiting ghosting artefacts after the removal of its first three singular values, compared to the slow dot, which is already splitting. The effect of flickering does not monotonously increase with the threshold but it is a balance between the residual energy and temporal variation of the residual singular vectors and may be less apparent when removing a higher number of singular values. Removing the first singular values coincides with stronger flickering but additional removal will reduce overall energy, thus also the amplitude of oscillation (Fig. \ref{fig01} e). In a real case ghosting and splitting will occur most likely with slow flow and where there is a change in the direction of flow (Fig. \ref{fig08}). It made no difference if the original dataset was complex radio frequency data, log compressed data or simple image data. This illustrates that the artefacts are caused by the SVD filter and are independent of the type of data that is being analysed. 

The results from the simulation and experiments each show different expressions of the flashing, ghosting and splitting artefacts. Flashing was the most common artefact when using SVD as a clutter filter. Remarkably, the temporal and spatial changes in local or global intensity are not present in the original data. They do not represent physiological information but might be caused by them e.g. cyclic expansion of the aorta as shown in Fig. \ref{fig07}. This is important to notice to avoid misinterpretation of data e.g. the changes in intensity of a microbubble in a filtered dataset could be falsely mistaken as its resonance response to the ultrasound transmission which in reality are on different time scales. The ghosting and splitting artefacts can most prominently be seen when signal is strong and sparse, e.g. when filtering contrast enhanced B-mode data as shown in Fig. \ref{fig04} and \ref{fig08}. The ability of SVD as a clutter filter depends on additional imaging parameters such as frame rate, stack size, field of view or tissue motion. Artefact mitigation strategies e.g. moving window SVD or block-wise SVD \cite{Song2016} can reduce artefacts but may have undesirable effects on the final image if other parameters are fixed. It is noteworthy that the artefacts observed in contrast enhanced imaging (all three) are different to those observed in contrast-agent free acquisitions (mainly flashing). Both imperfect separation of tissue and flow and over- or underestimation of the tissue domain can cause these artefacts. If the initial flow signal is very strong e.g. the high signal intensity of microbubbles in contrast enhanced imaging the use of SVD as a clutter filter can introduce artefacts unnecessarily. Other filtering techniques might be more favourable. Using the SVD clutter filter to reduce noise in a dataset can change the shape of the final image as shown in Fig. \ref{fig04}. The correct estimation of the clutter domain is key and is not trivial. Finally, it must be stated that SVD filtering can introduce artefacts as any partial projection of the original data can, but to date SVD based filtering remains one of the best clutter filters that exists which explains its rightful use.  

 The impact of the introduced artefacts on velocity gradient measurements for near wall flow or on the number of localization for super-resolution imaging was not assessed. However, the change in local image intensity and distribution as depicted in Fig. \ref{fig07}, Fig. \ref{fig04} and Fig. \ref{fig08} suggests that the SVD filter can alter these metrics \cite{Desailly2016, Geunyong2014}. In super localisation image process, it is assumed that each individual microbubble exhibits similar scattering properties during the acquisition time. The signal from a single microbubble is identified across multiple frames and distinguished from signals originating from noise and multiple microbubbles \cite{Jeffries2014}. Ghosting and signal splitting due to the SVD filter (Fig. \ref{fig08}) changes the shape of the signal and might lead to false identification. Similarly, the large and rapid changes in intensity can influence a cross-correlation based ultrasound image velocimetry algorithm (Fig. \ref{fig07}). The tapering of the flow region during slow flow (Fig. \ref{fig09}) and large changes in regional intensity in contrast-agent free acquisitions shows that the accurate measurement of near wall flow is limited at these times of the cardiac cycle \cite{Riemer2021}.
 
It has been argued that the flashing artefact created by the SVD clutter filter can be attributed to using a cut-off point that is too low \cite{Baranger2018}. However, the results in Fig. \ref{fig07} show that in certain cases with motion the flashing cannot be removed. After SVD clutter filtering, but not before, regions of both examples exhibited periodic spikes in intensity. In Fig. \ref{fig07} b these oscillations correlated with the cardiac cycle, the peak intensity occurring at peak systole, which is also the point where the maximum vessel motion occur. Regardless of the cut-off above singular value 12, the mean intensity of the reconstructed image changed at 3 Hz, corresponding to the frequency of motion. A higher cut-off point will result in less visible image fluctuations, but not eliminate the flashing, and will reduce the CTR. To reduce the flashing artefact, the acquisition could be divided in multiple spatial regions \cite{Song2016} and temporal sections of less variable flow velocity and tissue motion as shown in Fig. \ref{fig09} b. For microvascular flow the velocities are usually more constant. However, large scale motion due to breathing, gut movement or twitching and the relatively long acquisition times can have a similar effect (Fig. \ref{fig07} c and d). Tissue motion can be challenging to correct but estimating motion can reduce the ghosting and splitting artefact (Fig. \ref{fig08}). MoCo also changed the relative magnitude and the sub-population of microbubbles that are suppressed. 

A limitation of this study is the curated selection of datasets. Artefacts are not always visible when using SVD. The given examples are catered to maximize the visual impact of the artefact for illustration purposes. However, in real-life the misuse of SVD as a clutter filter can commonly be seen. More than most filters SVD as a clutter filter requires an adequate imaging pipeline e.g. from a high frame rate to post-acquisition MoCo. Regardless of the settings SVD-based clutter filtering increases the SNR, therefore the presented artefacts may be easily overlooked or misinterpreted by the user. 

Flow velocity (Fig. \ref{fig01} and \ref{fig09}), distribution of scatters (Fig. \ref{fig02}, \ref{fig04} and \ref{fig08}), the choice of imaging and filter parameters (Fig. \ref{fig05}) the filter cut-off points (Fig. \ref{fig04}) and motion (Fig. \ref{fig07}) all influence the generation and the extent of the SVD artefacts. Ghosting and splitting artefacts will mostly occur when using contrast enhanced data. While the spatiotemporal characteristics of the flow such as velocity, pulsatility and microbubble distribution cannot be changed, the imaging parameters and filter settings can be altered. Fig. \ref{fig04} shows that not all changes in magnitude and shape of the signal can be classed as an artefact but rather the effect of SVD as a clutter filter. Small changes in threshold can lead to large differences in the appearance of the final image. Finally, a weak general statement can be made about mitigating SVD clutter artefacts. A low cut-off reduces the risk of ghosting and target splitting, but in exchange can increase the flashing artefact. A high cut-off threshold reduces the appearance of flashing but comes at the cost of a loss in SNR and may cause ghosting and splitting. Filter artefacts are not unique to SVD based clutter rejection; but rather that the choice of a method for clutter rejection should consider the designated application, the benefit or renunciation of contrast agents, desired frame rate and the spatiotemporal properties of the imaging region. 

\section{Conclusion}
We demonstrated flashing, ghosting and splitting artefacts caused by SVD-based clutter filtering and investigated the underlying causes and influencing factors. Nevertheless, SVD as a clutter filter remains a powerful technique and its potential benefits have not been fully realised. 

\appendices

\section*{Acknowledgment}
This work was supported in part by the Engineering and Physical Sciences Research Council (EPSRC) under Grant EP/T008970/1, and EP/N026942/1, the EPSRC Impact Acceleration Account under Grant EP/R511547/1, the National Institute for Health Research i4i under Grant NIHR200972, and the British Heart Foundation under Grant PG/18/48/33832.

\end{document}